# The Chromospheric Solar Millimeter-wave Cavity; a Common Property in the Semi-empirical Models

*V. De la Luz[1,2], M. Chavez[2], and E. Bertone [2]

1 Universidad Autónoma Metropolitana Iztapalapa, Departamento de Física, A.P. 55-534, C.P. 09340, D.F., México.
2 Instituto Nacional de Astrofisica, Optica y Electronica, Tonantzintla, Puebla, Mexico, Apdo. Postal 51 y 216, 72000

**Abstract**

The semi-empirical models of the solar chromosphere are useful in the study of the solar radio emission at millimeter - infrared wavelengths. However, current models do not reproduce the observations of the quiet sun. In this work we present a theoretical study of the radiative transfer equation for four semi-empirical models at these wavelengths. We found that the Chromospheric Solar Milimeter-wave Cavity (CSMC), a region where the atmosphere becomes locally optically thin at millimeter wavelengths, is present in the semi-empirical models under study. We conclude that the CSMC is a general property of the solar chromosphere where the semi-empirical models shows temperature minimum.

Los modelos semi-empíricos de la cromosfera solar son herramientas importantes en el estudio de la radio emisión solar a longitudes de onda milimétricas-infrarrojas. Sin embargo, los modelos actuales siguen sin reproducir las observaciones del sol quieto. En este trabajo, presentamos un estudio teórico de la ecuación de transferencia radiativa para cuatro modelos semi-empíricos a estas longitudes de onda. Encontramos que la Cavidad Cromosferica Solar a Longitudes de Onda Milimétricas (CSMC), una región donde la atmósfera se vuelve opticamente delgada localmente a longitudes de onda milimétricas, esta presente en los modelos semi-empíricos bajo estudio. Concluimos que la CSMC es una propiedad general de la cromosfera solar donde los modelos cromosfericos muestran un mínimo de temperatura.

Keywords: Sun: chromosphere — Sun: radio radiation — Sun: infrared — methods: numerical — radiative transfer — stars: chromospheres

## 1. Introduction

In 1902, the first theoretical computation of the stellar radio emission using a black body at 5700 K showed that the flux that emerges of the solar surface was almost impossible, at that eppoch, to observe from ground based observations due to low computed flux and the sensitivy of the instruments, which discouraged the first attempts to observe the solar radio emission (Nordmann, 1905).

Fourty years after the Planck theoretical computations, the first radio observation of the quiet sun was confirmed (Reber, 1944; Martyn, 1946). Observations (Pawsey and Yabsley, 1949; Zirin, Baumert, and Hurford, 1991; Vourlidas et al., 2010), and the theoretical models where subsequently improved from two steps models (cold-hot) to sophisticad hydrostatic models (Smerd, 1950; van de Hulst, 1953; Allen, 1963; Ahmad and Kundu, 1981; Vernazza, Avrett, and Loeser, 1981; Fontenla et al., 2011).

The hydrostatic semi-empirical models showed that the stratification of the chromosphere could explain the continuum in the millimeter-infrared spectral region. The UV emission becomes the major point of reference to calibrate the semi-empirical models while the radio continuum was used only to test the auto consistence of the models (Vernazza, Avrett, and Loeser, 1981).

The semi-empirical models of the quiet sun chromosphere have two hypotheses: the magnetic field at these scales have no effect in the convective flux and the vertical scale is lower than the horizontal scale (Fontenla et al., 2006). The results of these hypotheses is a stratified plain-parallel atmosphere in hydrostatic equilibrium. In the UV-visible region have provided good approximations, however differences between the synthetic spectra and the observations, specially those associated at altitudes around the temperature minimum of the solar chromosphere, can be observed at millimeter and infrared regions (Landi and Chiuderi Drago, 2003).

Regardless of the theoretical approaches, we now know that the chromosphere is a very reach region of the solar atmosphere where the magnetic field at micro scales plays an important role in the morphology of this layer (Vourlidas et al., 2010). However, the infrastructure required to observe the micro structure at radio frequencies is still beyond current observational capabilities.

In this work, we study the radiative transfer equation specially at heights associated with the temperature minimum of the solar chromosphere to reproduce the Chromospheric Solar Millimeter-wave Cavity (CSMC) found in De la Luz, Raulin, and Lara (2013) using 4 different semi-empirical models as input: VALC from Vernazza, Avrett, and Loeser (1981), SRPM305 from Fontenla et al. (2006), the cold [1000A], and hot [1008Q] models from Fontenla et al. (2011). These models have the general characteristic that they incorporate a temperature minimum region whose millimeter emission is used by the authors as a test of the auto consistency of their models.

The goal of this work is to explore if CSMC is a general property of solar atmospheric models that include a temperature minimum region.

## 2. The Semi-empirical Models

In Figure 1 we plot the temperature profile for the VALC, SRPM305, 1000A, and 1008Q models. We can observe the temperature minimum region between 100 and 1000 km over the photosphere. The temperature profiles show a decrement of the temperature that comes from the photosphere until reaches to the minimum value of temperature, then the gradient inverts and the temperature increments, a plateau of temperature of around 1000 km is presented in all the models, finally the temperature grows until coronal temperatures of around 1e6 K.

For the density (figure 2), the models shows a exponential decrease starting at photospheric altitudes until altitudes around the value of minimum temperature. The density profiles depend directly of the temperature profile: for lower temperatures the density is also lower. Finally, at high altitudes with respect the photosphere, the density profile is correlated with the increase of the temperature profile towards coronal temperatures. In this region, the density drops two orders of magnitude to coronal density values ($\rho \approx 1e7$).

## 3. Computations

We used the code PakalMPI (De la Luz, Lara, and Raulin, 2011) to solve the radiative transfer equation. The code is written in C/MPI with GNU/GPL License. PakalMPI take as input the hydrogen density, temperature, and metallicity radial profiles; computes the ion abundances in LTE for 18 atoms and the NLTE abundances for H, H-, and electrons. Then, the code calculates the ray path and solve, using integrations step controlled by an intelligent algorithm, the radiative transfer equations using three opacity functions for Bremsstrahlung, H− , and Inverse Bremsstrahlung. Since the opacity functions depend on the number of ion and electrons, the densities of these charged particles are also computed by PakalMPI. Finally, the brightness temperature, the optical depth and the opacities are printed step by step in altitude at each frequency. This information is used for our analysis to compute the local emissivity:

$$E_l = 1 - \exp(-\tau_{local})$$

where $\tau_{local}$ is the local optical depth. The $E_l$ parameter shows the capability of the atmosphere to generate radiation. When $E_l \approx 0$ the atmosphere is transparent (optically thin) and if $E_l \approx 1$ the atmosphere is optically thick. We use the $E_l$ value as a diagnostic of the radiative transfer in the solar chromosphere (De la Luz, Raulin, and Lara, 2013).

## 4. Results

Figure 3 shows the local emissivity ($E_l$) for the four semi-empirical models under study plotting height vs frequency over the photosphere and in colors the $E_l$ parameter or equivalently, the altitude where the emission is generated. For the four semi-empirical models the region where the atmosphere is locally optically thick is presented as a peak around 1000 km over the photosphere. Below this peak, a region where locally the atmosphere is transparent is also presented. This region is what we have called the CSMC. We found that the CSMC is present in all models. In Figure 4, we show that the depth of the cavity, as a function of frequency, reaches lower frequencies for lower temperature minimum. In Figure 5 we show the relation between the cavity and the density profile. The continuous line show the relation between the height over the photosphere where the peak of the CSMC is maximum in frequency and their density in the semi-empirical model at the same altitude. A dependence between frequency and density is not evident for the case of the peak of the CSMC.

## 5. Conclusions

We found the CSMC in the four semi-empirical models under study. Figures 4 and 5 show that the temperature plays an important role in the depth in frequency of the cavity, however a clear relation between the density and the peak of the cavity is not clearly shown. The peak of the cavity is important because it is the responsible of the morphology of the spectrum at sub-millimeter and infrared wavelengths, and its characterization is fundamental to develop more realistic models that ameliorate the discrepancies between theory and observations. We have found that the CSMC is present in the four semi-empirical models we analyzed. This result indicates that CSMC might be a general property of physical systems characterized by having a temperature minimum region.

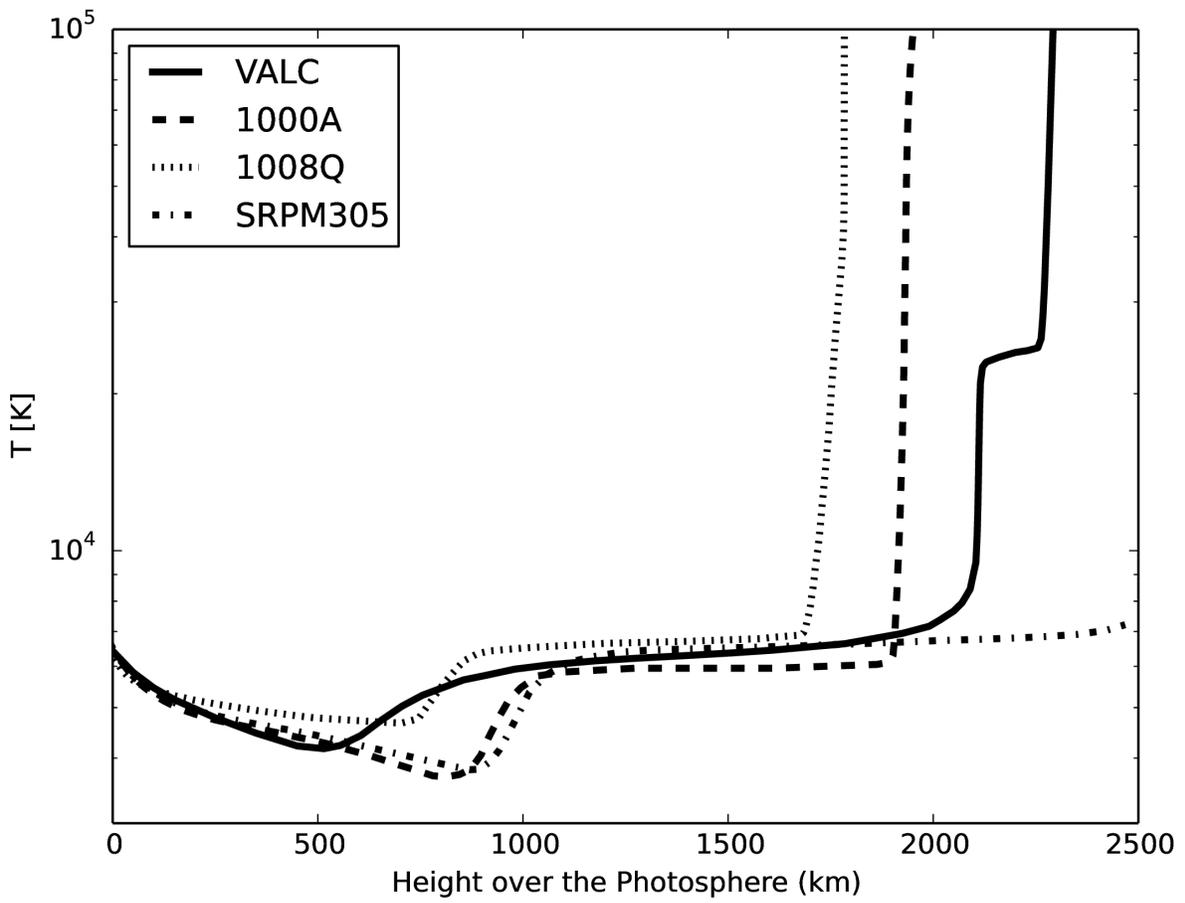

Figure 1. Temperature profiles for semi-empirical models.

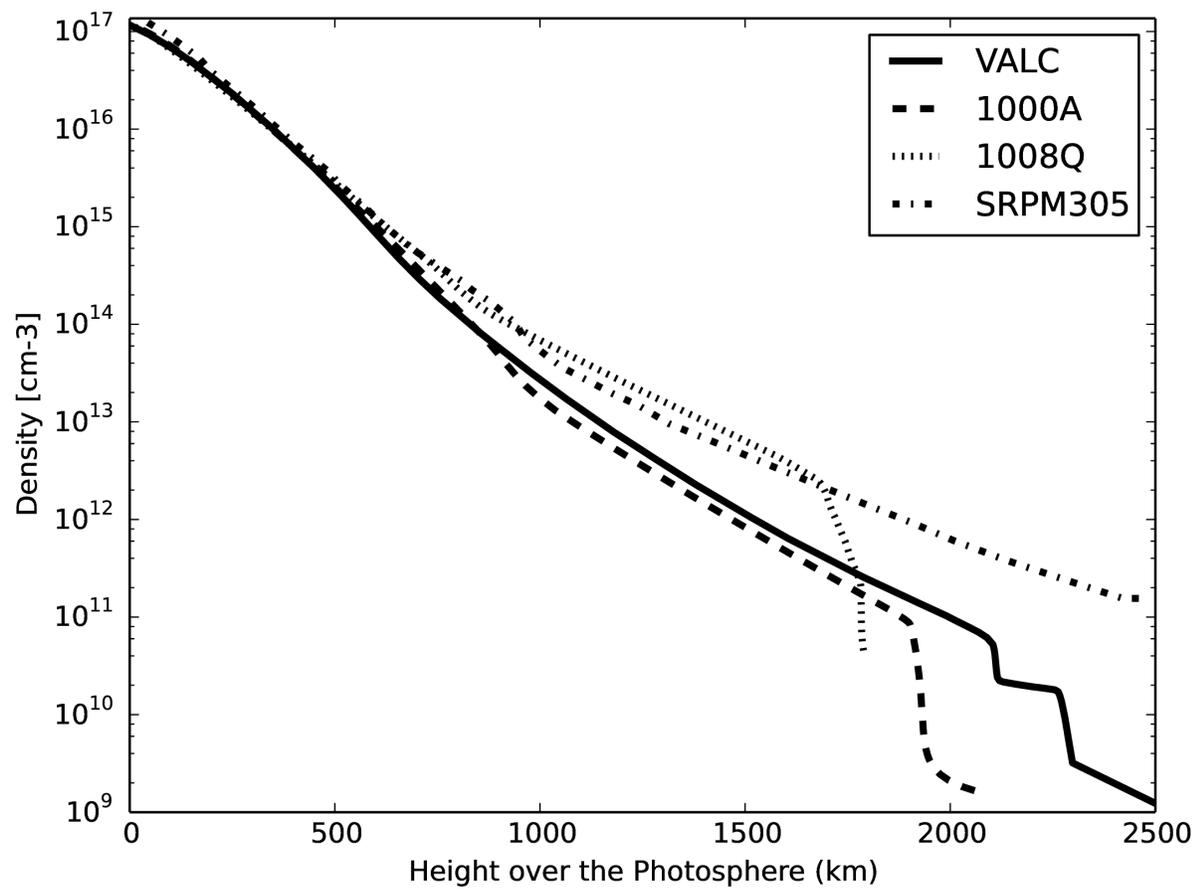

Figure 2. Density profiles for semi-empirical models.

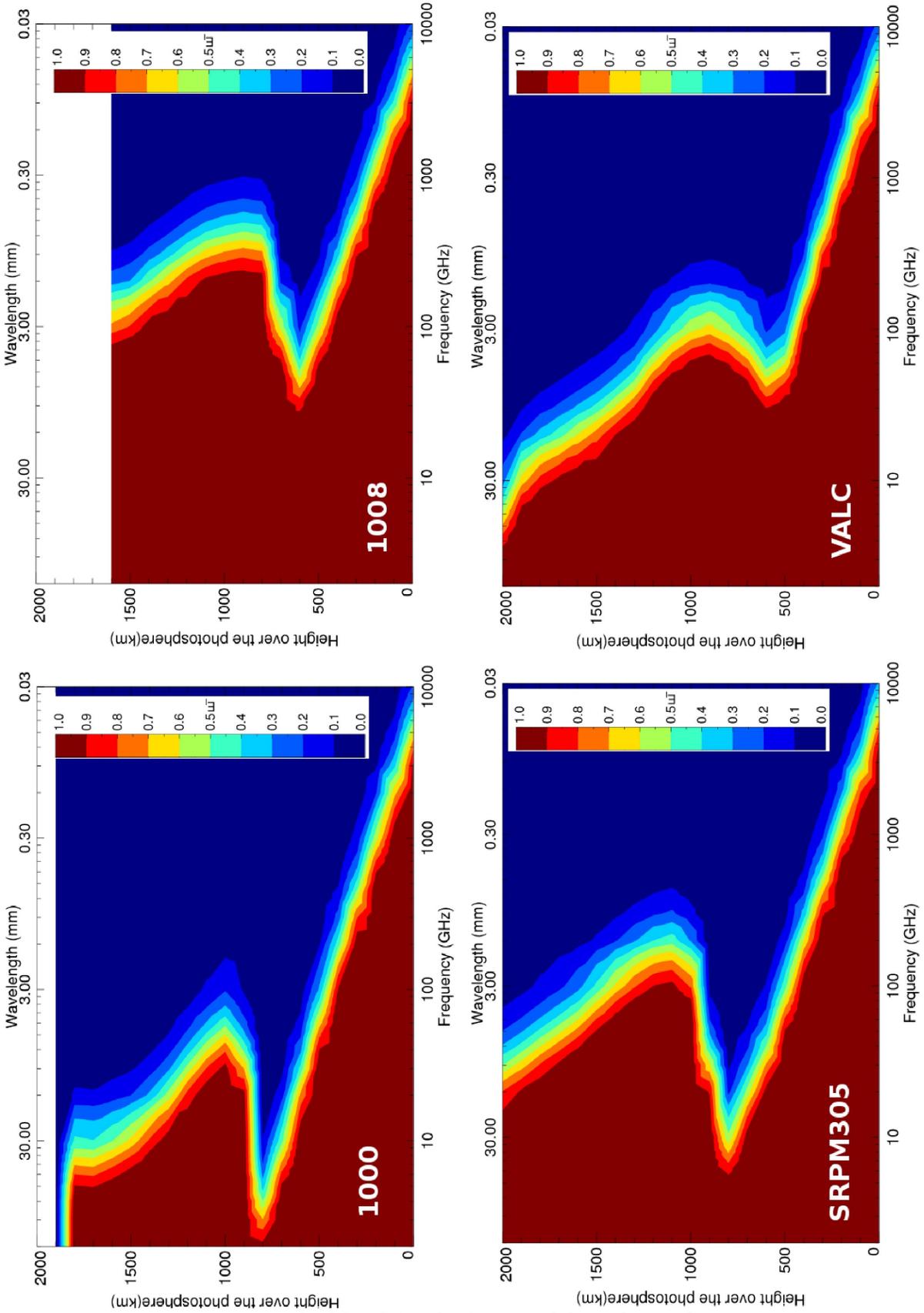

Figure 3. CSMC for the four models under study.

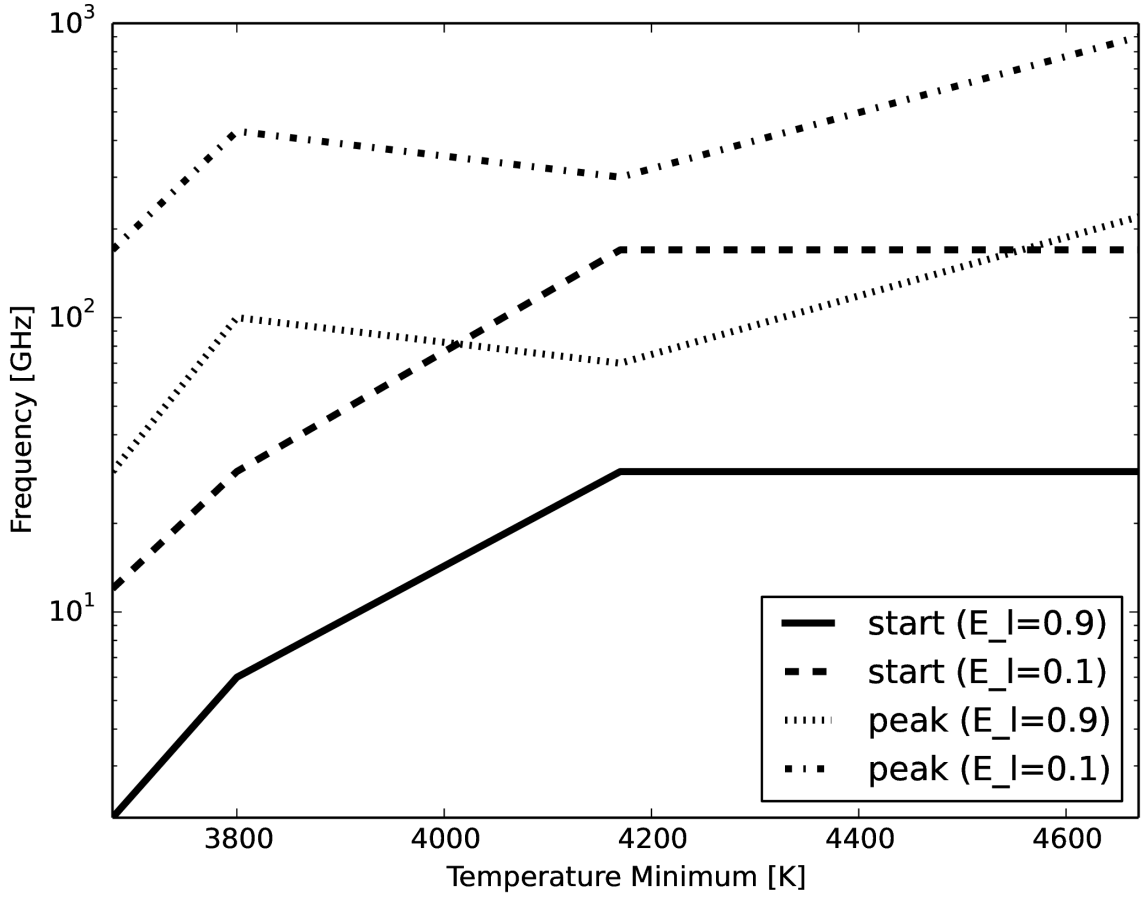

Figure 4. Relation between the value of the temperature minimum and the CSMC for the start of the cavity (for two local emissivity thresholds, 0.9 and 0.1) and the peak of the cavity (also for $E_l = 0.9$ and 0.1). We have considered the following temperature minima for the four models under analysis: 1000A ~ 3680K; SRPM305 ~ 3800K; VALC ~ 4170K; 1008Q ~4670K.

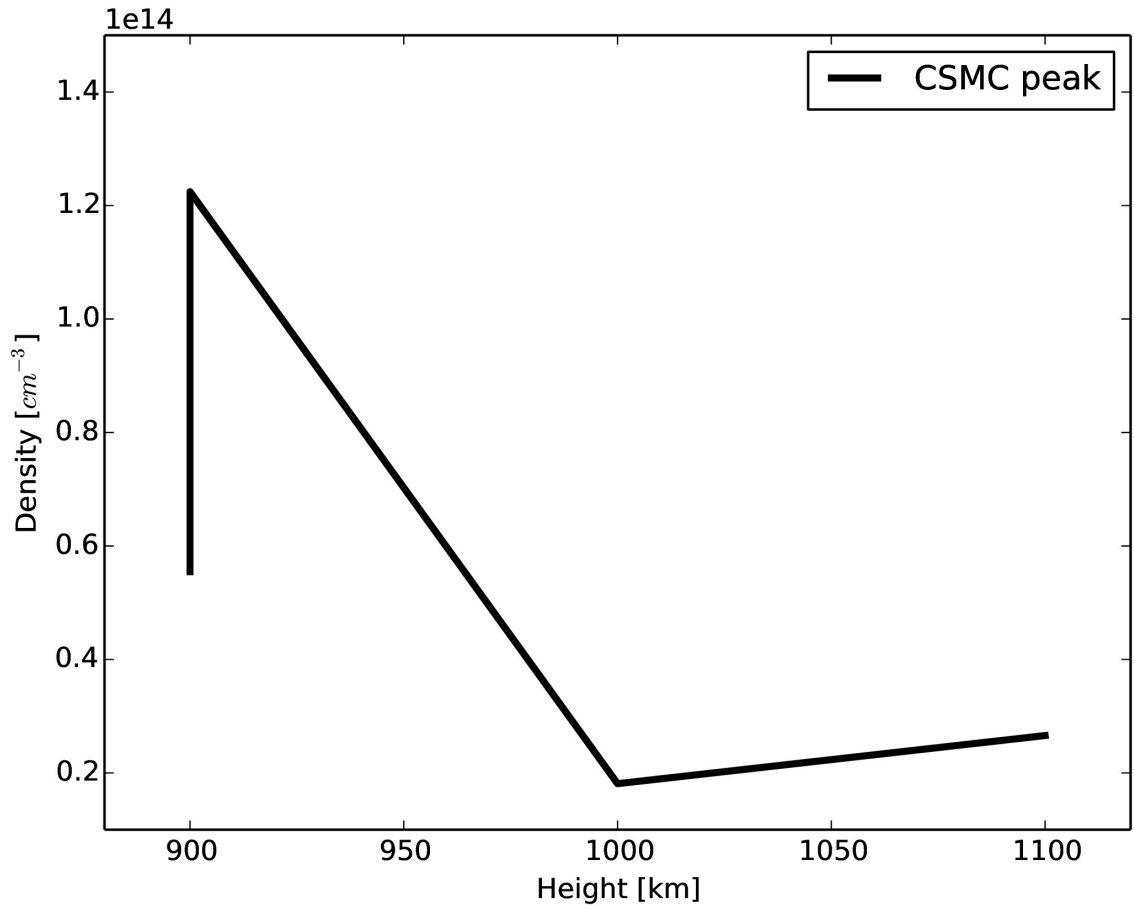

Figure 5. Relation between the height over the photosphere and density at the altitude where the peak of the cavity is maximum ($E_l = 0.1$).